\documentclass[numsec,webpdf,modern,medium,namedate]{oup-authoring-template}

\usepackage{setspace}
\usepackage[margin=1in]{geometry}
\usepackage{titlesec}
\titleformat{\section}{\large\bfseries}{\thesection}{1em}{}
\titleformat{\subsection}{\normalsize\bfseries}{\thesubsection}{1em}{}

\usepackage{tikz}
\usetikzlibrary{positioning,calc}
\usepackage[export]{adjustbox}

\onecolumn

\graphicspath{{figures/}}

\usepackage{booktabs}
\theoremstyle{plain}

\theoremstyle{plain}

\theoremstyle{thmstylethree}

\usepackage{booktabs}
\usepackage{siunitx}
\usepackage[fontsize=12pt]{fontsize}
\usepackage{bbm}
\usepackage{tabularx}
\usepackage[export]{adjustbox}

\begin{document}

\journaltitle{Journals of the Royal Statistical Society}
\DOI{DOI HERE}
\copyrightyear{XXXX}
\pubyear{XXXX}
\access{Advance Access Publication Date: Day Month Year}
\appnotes{Original article}

\firstpage{1}

\title[Bayesian Downscaling]{Bayesian Downscaling of Satellite-Derived Solar-Induced Chlorophyll Fluorescence}

\author[1,$\ast$]{Russell Goebel}
\author[2]{Leeza Moldavchuk}
\author[3]{Taylor S. Jones}
\author[3]{Jonathan Dooley}
\author[2]{Lucy R. Hutyra}
\author[1]{Luis Carvalho}

\authormark{Goebel et al.}
\address[1]{\orgdiv{Mathematics \& Statistics}, \orgname{Boston University}, 
  \orgaddress{\street{775 Commonwealth Avenue, Boston}, \postcode{02215}, \state{MA}, \country{United States}}}
\address[2]{\orgdiv{Earth \& Environment}, \orgname{Boston University}, 
  \orgaddress{\street{775 Commonwealth Avenue, Boston}, \postcode{02215}, \state{MA}, \country{United States}}}
\address[3]{\orgdiv{The Earth Commons}, \orgname{Georgetown University}, 
  \orgaddress{\street{3700 O St., NW}, \postcode{20057}, \state{Washington DC}, \country{United States}}}

\corresp[$\ast$]{Address for correspondence. Russell Goebel, Boston University, Boston, 02215, United States. \href{Email:email-id.com}{russjgoebel@gmail.com}}

\received{Date}{0}{Year}
\revised{Date}{0}{Year}
\accepted{Date}{0}{Year}

\abstract{
We present an extendable Bayesian framework for spatial downscaling of satellite-derived measurements with coarse, irregular footprints. Motivated by measurements of solar-induced chlorophyll fluorescence (SIF) from the earth-observing OCO-2 satellite, we model coarse-resolution observations as noisy, weighted averages over an underlying fine-resolution spatial field. The latent process is defined over a regular grid and regularized using a Simultaneous Autoregressive (SAR) prior, yielding a sparse precision structure that enforces spatial smoothness. This formulation enables efficient posterior inference including uncertainty quantification and naturally accommodates land cover information through modified neighborhood structures or covariates. To assess accuracy, we conduct a semi-synthetic validation using albedo data derived from Sentinel-2 satellite surface reflectance, finding strong correspondence between posterior means and the synthetic generating field ($R^2 = 0.72$ when including the proportion of water as a covariate and 0.57 without any covariates).  We apply our method to a high-density OCO-2 target mode SIF observations over Boston, demonstrating that it produces fine-resolution posterior maps that reflect heterogeneity in urban environments and align with greenness, with a moderate relationship with the Normalized Difference Vegetation Index (NDVI) (R$^2$ = 0.417 for pixels with 20+ overlapping soundings, 0.323 overall). Our approach achieves high-resolution recovery of satellite measurements with principled uncertainty quantification.
}

\maketitle

\section{Introduction}

Solar-induced chlorophyll fluorescence (SIF) has emerged as a promising remotely sensed indicator of gross primary productivity (GPP), a quantity that cannot be directly measured at the scales needed for global climate and carbon cycle modeling \citep{porcar2021chlorophyll,MOHAMMED2019111177}. Emitted in the red and far-red wavelengths as chlorophyll dissipates absorbed solar energy \citep{GROSSMANN2018311}, SIF is strongly associated with GPP across diverse biomes \citep{li2018solar}. Vegetation indices such as the enhanced vegetation index (EVI) and normalized difference vegetation index (NDVI) are widely used as proxies for GPP; however, these primarily reflect photosynthetic capacity rather than real-time activity. In contrast, SIF is more directly linked to instantaneous photosynthetic processes. This growing interest in SIF has been accompanied by increasing attention to downscaling (disaggregation or oversampling) of remotely observed SIF measurements, as many scientific and policy-relevant questions---particularly those concerning heterogeneous landscapes such as cities, agricultural mosaics, and fragmented forests---require spatial resolution finer than that provided by existing satellite products.

\begin{figure}
    \centering
    \includegraphics[width=0.45\linewidth]{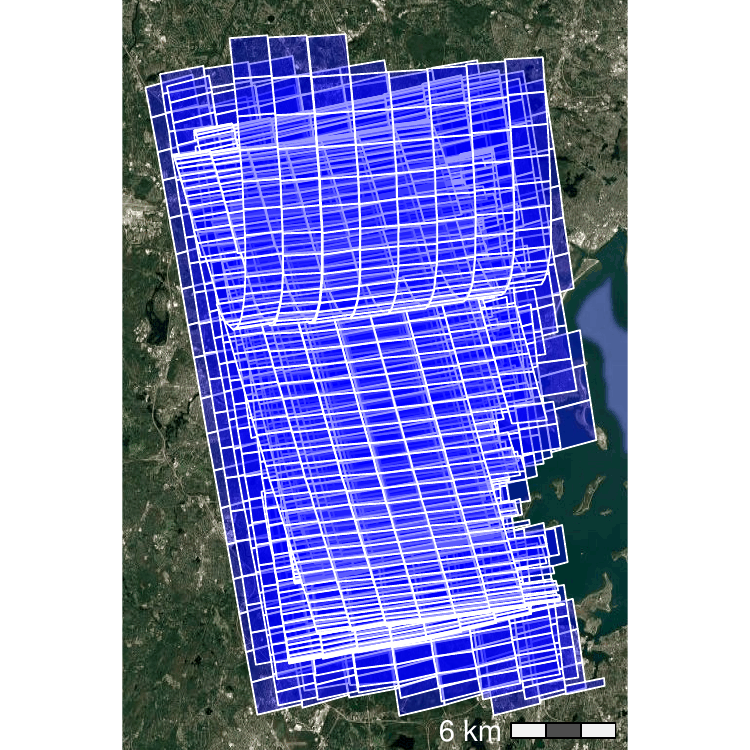}
    \includegraphics[width=0.45\linewidth]{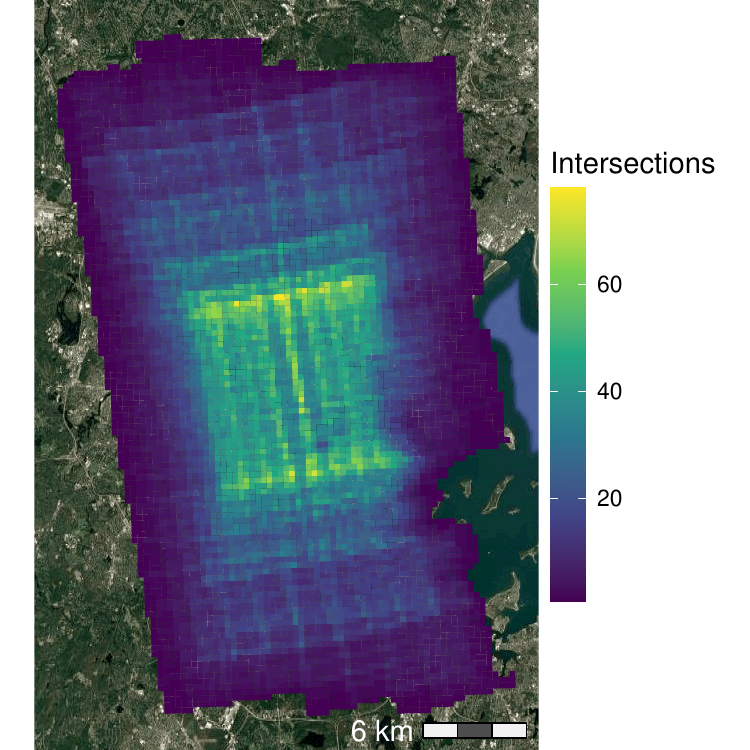}
    \caption{OCO-2 sounding geometries (left) and number of overlapping soundings (right) on a 330 meter resolution grid over Boston as measured on June 14, 2022. This is the same grid used in the rest of the analyses (e.g, Figures \ref{fig:combined_downscaling_plots} and \ref{fig:downscaling_plots_ndvi_se}.)}
    \label{fig:oco_overlap}
\end{figure}

In parallel with methodological developments in downscaling and oversampling, satellite observation strategies have also evolved to provide denser spatial sampling. In particular, the Orbiting Carbon Observatory-2 (OCO-2) satellite offers a ``target mode" acquisition strategy in which the instrument repeatedly scans a fixed region for several minutes, producing thousands of partially overlapping soundings over areas approximately 15--20 km in extent \citep{eldering2017orbiting}. An example of the retrieval density used in this paper can be seen in Figure \ref{fig:oco_overlap}. Each sounding corresponds to a footprint on the order of 1--2 km and represents a spatially averaged measurement rather than a point observation. This dense, overlapping sampling regime --- with thousands of partially overlapping kilometer-scale footprints over a 15--20\,km region --- 
creates both an opportunity and a statistical challenge: the overlapping 
geometry carries sub-footprint information, but recovering it requires 
a model that explicitly accounts for the areal nature of each 
measurement. The task of translating these measurements into a gridded product can be seen as transitioning between satellite data processing levels: standardized across agencies including NASA and NOAA \citep{weaver2014processing}, these levels reflect increasing degrees of preprocessing, where the overlapping observations are an example of a level-2 product, and a desired regular gridded product an example of level-3. 

Existing SIF downscaling approaches — regression-based (including machine learning), geometry-informed oversampling (e.g \citealt{turner2020double} or \citealt{sun2018physics}), and geostatistical (e.g, kriging), sometimes combined -- each address part of the problem but none simultaneously treats observations as areal integrals, scales to thousands of soundings, and provides calibrated uncertainty. The first tradition, regression-based downscaling, learns 
$f$ such that $y_c \approx f(X_c)$ and predicting $\hat{y}_f = f(X_f)$, where 
$y_c$ and $y_f$ denote SIF at the coarse and fine scales and $X_c$, $X_f$ are 
auxiliary covariates at the respective scales. An early and influential example is \citet{DUVEILLER201672}, who 
downscaled GOME-2 SIF from $0.5^\circ$ to $0.05^\circ$ using a 
locally-calibrated light-use efficiency model with NDVI, evapotranspiration, 
and land surface temperature as covariates. Machine learning 
methods often similarly predict $\hat{y}_f = f(X_f)$, including the neural network used by 
\citet{zhang2018global} to produce CSIF, a $0.05^\circ$ product reconstructed 
from MODIS reflectance, and the cubist regression trees of \citet{li2019global} 
used to produce GOSIF, a $0.05^\circ$ product combining MODIS with meteorological 
reanalysis. A related but distinct approach is taken by 
\citet{gensheimer2022convolutional} in SIFnet, a 500\,m product based on a 
convolutional neural network that takes coarse-resolution TROPOMI SIF directly 
as an input alongside fine-resolution auxiliary data, i.e., $\hat{y}_f = 
f(y_c, X_f)$, learned by training on artificially degraded SIF observations. These approaches share a common limitation: they are not applied to the raw overlapping footprints, assume the covariate-SIF relationship is preserved across scales, and provide no calibrated uncertainty over the fine-resolution field.

These approaches operate on SIF observations that have already been aggregated to a regular coarse grid, sidestepping the irregular geometry of the underlying satellite footprints. However, geometry-informed approaches for downscaling SIF are present in the literature, with a notable example being \cite{turner2020double} whose oversampling approach redistributes coarse signal according to footprint overlap. These methods have gained considerable 
traction --- \citet{turner2020double} in particular produces a widely-used 500\,m 
SIF product over CONUS---but are not grounded in a probabilistic model and 
provide no uncertainty estimates.

A principled alternative---change-of-support kriging---requires evaluating a covariance matrix whose entries depend on integrals over pairs of spatial supports. In our setting this amounts to nearly 50 million double integrals before any matrix factorization — a cost that at target-mode observation densities dwarfs the subsequent cubic factorization that is well documented as the bottleneck in large spatial data problems
\citep{banerjee2017high, heaton2019case}. Scalable spatial 
models---including nearest-neighbor Gaussian processes, multi-fidelity 
formulations, and related precision-based constructions---have 
substantially extended the range of feasible spatial inference for point-data
\citep[e.g.,][]{cheng2024recursive, ma2022multifidelity, 
calle2025generative}, but are typically only formulated for observations 
indexed at locations rather than as overlapping spatial integrals 
over irregular supports. The dominant geostatistical framework in 
this literature, area-to-point kriging \citep[ATPK;][]{park2019overview}, 
treats the target as a point process despite the areal nature of both 
the source observations and the desired gridded product. A notable 
SIF-specific example is \citet{jacobson2023spatial}, who use cokriging 
of OCO-2 SIF and XCO$_2$ to produce a gridded Level-3 product with 
formal uncertainty quantification; however, to remain computationally 
tractable, the irregular observations are averaged to the grid prior 
to model fitting and treated as point observations, with inference 
carried out over a 300-pixel moving window. More generally, treating irregular areal observations as point measurements at their centroids ignores within-footprint variance, understating prediction uncertainty and leading to undercoverage of the true fine-resolution field \citep{goovaerts2008kriging}. These geostatistical approaches thus face a trilemma: full change-of-support kriging handles the geometry correctly but does not scale; scalable alternatives are practical but are formulated for point-indexed observations; and approaches that aggregate irregular footprints to a grid prior to fitting discard sub-footprint information and understate uncertainty.

To our knowledge, no existing SIF downscaling framework simultaneously models observations as areal integrals accommodating irregular footprints and produces a calibrated posterior distribution over the fine-resolution field at the scale of target-mode acquisitions. The framework proposed here addresses this gap through a sparse, consistency-preserving discretization of the continuous change-of-support integral that reduces inference over thousands of irregular overlapping footprints to standard sparse Bayesian linear regression — without evaluating the pairwise spatial integrals that make full area-to-area kriging computationally prohibitive. This discretization is the key ingredient that simultaneously enables calibrated posterior uncertainty, accommodation of auxiliary covariates, and scaling to the observation densities characteristic of target-mode acquisitions — a combination no existing framework achieves. The latent field is regularized by a simultaneous autoregressive (SAR) prior, and spatial confounding between the covariate and spatial residual is addressed via restricted spatial regression in our SIF application, where the sharp land-water boundary would otherwise produce physically implausible coefficient estimates.

In addition to the SIF measurements from OCO-2, we generate a semi-synthetic albedo dataset by adding random noise to a Sentinel-2 derived albedo product to validate the approach. We assess robustness to hyperparameter selection and inclusion of a covariate.  Because albedo exhibits high spatial heterogeneity in urban environments, it provides a meaningful benchmark for evaluating how well fine-scale structure is recovered under a known reference.

\section{Downscaling Methods}


\begin{table}[t]
\small     
\caption{Summary of notation used throughout the paper.}
\label{tab:notation}
\begin{tabular}{lll}
\toprule
Symbol & Type & Description \\
\midrule
\multicolumn{3}{l}{\textit{Domains and indices}} \\
$\mathcal{D}$ & set & spatial domain \\
$n$ & scalar & number of observations \\
$m$ & scalar & number of latent grid cells \\
$D^{(o)}_i$ & set & support of the $i$-th observation footprint \\
$D^{(\ell)}_j$ & set & support of the $j$-th latent grid cell \\
\midrule
\multicolumn{3}{l}{\textit{Observed and latent variables}} \\
$v(s)$ & function & underlying spatial process at location $s$ \\
$y^{(o)}_i$ & scalar & $i$-th observed measurement ($n \times 1$ vector $y^{(o)}$) \\
$y^{(\ell)}_j$ & scalar & mean of the $j$-th latent grid cell ($m \times 1$ vector $y^{(\ell)}$) \\
$r(s)$ & function & zero-mean residual spatial process \\
$r^{(\ell)}$ & vector & discretized residual at latent resolution ($m \times 1$) \\
$\varepsilon_i$ & scalar & measurement error for observation $i$ \\
$\Sigma_\varepsilon$ & matrix & measurement error covariance ($n \times n$) \\
\midrule
\multicolumn{3}{l}{\textit{Covariates}} \\
$x(s)$ & vector & $p \times 1$ covariate vector at location $s$ \\
$X^{(o)}$ & matrix & aggregated covariates at observation scale ($n \times p$) \\
$X^{(\ell)}$ & matrix & covariates at latent grid resolution ($m \times p$) \\
$\beta$ & vector & covariate coefficients ($p \times 1$) \\
\midrule
\multicolumn{3}{l}{\textit{Aggregation operator}} \\
$g_i(s)$ & function & sensor footprint response function for observation $i$ \\
$A$ & matrix & aggregation (change-of-support) matrix ($n \times m$),
              $A_{ij} = \frac{|D^{(o)}_i \cap D^{(\ell)}_j|}{|D^{(o)}_i|}\bar{g}_{ij}$ \\
\midrule
\multicolumn{3}{l}{\textit{Prior and posterior}} \\
$W$ & matrix & adjacency matrix for SAR prior ($m \times m$) \\
$\widetilde{W}$ & matrix & row-normalised adjacency, $\widetilde{W}_{ij} = 1/n_i$ for neighbours \\
$\rho$ & scalar & SAR spatial range parameter, $W = \rho\widetilde{W}$ \\
$\tau$ & scalar & SAR smoothing (precision scaling) parameter \\
$Q$ & matrix & SAR structure matrix, $(I - W)^\top(I - W)$ ($m \times m$) \\
$\Lambda_\beta$ & matrix & prior precision for $\beta$ ($p \times p$) \\
$\Lambda_{\beta,r}$ & matrix & combined prior precision, $\mathrm{blockdiag}(\Lambda_\beta, \tau Q)$
                               ($(p+m) \times (p+m)$) \\
$\sigma^2$ & scalar & observational variance \\
$\gamma$ & vector & concatenated parameter vector $[\beta^\top, r^{(\ell)\top}]^\top$ ($(p+m) \times 1$) \\
$\Phi$ & matrix & concatenated design matrix $[X^{(o)}\ A]$ ($n \times (p + m)$) \\
$\Lambda_\gamma$ & matrix & posterior precision of $\gamma$ ($(p+m) \times (p+m)$) \\
$\mu_\gamma$ & vector & posterior mean of $\gamma$ ($(p+m) \times 1$) \\
\midrule
\multicolumn{3}{l}{\textit{RSR constraint}} \\
$C$ & matrix & RSR constraint matrix, $C = X^{(o)\top}A$ (or QR-reduced form) \\
$B$ & matrix & basis for nullspace of $C$, used to reparametrise $r^{(\ell)}$ \\
\bottomrule
\end{tabular}
\end{table}

The spatial domain $D$ is partitioned into $n$ observation regions $D_i^{(o)}$ and $m$ latent prediction regions $D_j^{\ell}$. Observed values are $y_i^{(o)}$, latent means are $y_j^{(\ell)}$, and the underlying spatial process is $v(s)$.  A summary of all notation used throughout this section and the rest of the paper is provided in Table~\ref{tab:notation}.

\subsection{Observation Model}

First, we model the underlying (unobserved) field with covariates as

\begin{align}
    v(s) = x(s)^\top \beta +r(s) \label{eq:v}
\end{align}
where $x(s) = \begin{bmatrix} x_1(s) & x_2(s) & \ldots x_p(s)\end{bmatrix}^\top$ are the covariate values and $r(s)$ a second-order process with mean 0 so that its variogram is well defined. No particular covariance family for $r$ is required.

We assume that the observed data $y_i^{(o)}$ can be written as
\begin{align}
y_i^{(o)} = \frac{1}{|D_i^{(o)}|} \int_{D_i^{(o)}} g_i(s) v(s)\, ds + \varepsilon_i, \label{eq:measurement}
\end{align}
 where $g_i$ can be thought of as the sensor's footprint response (e.g., a super-Gaussian distribution). In our application we use a uniform response $g_i(s) = 1$ within each footprint, consistent with the motivation of treating observations as areal averages rather than point measurements; sensitivity to this assumption is assessed in the Supplement (Section~6), where we find that misspecification primarily degrades uncertainty quantification rather than point prediction accuracy.
 Targets are fine--scale region means

\begin{align}
y_j^{(\ell)} &= \frac{1}{|D_j^{(\ell)}|} \int_{D_j^{(\ell)}} v(s) \, ds \notag\\
&= \frac{1}{|D_j^{(\ell)}|}\int_{D_j^{(\ell)}}x(s)^\top \beta d(s)+\frac{1}{|D_j^{(\ell)}|}\int_{D_j^{(\ell)}}r(s) d(s) \notag \\
&= x_j^{(\ell)\top} \beta + r_j^{(\ell)},
\end{align}

where $x_j^{(\ell)} = \frac{1}{|D_j^{(\ell)}|}\int_{D_j^{(\ell)}}x(s)ds$ and $r_j^{(\ell)} = \frac{1}{|D_j^{(\ell)}|}\int_{D_j^{(\ell)}}r(s) d(s).$

The objective is to estimate and model uncertainty of $y_j^{(\ell)}$ using $y_{i}^{(o)}$ and the covariate information.

\subsection{Discretization}\label{sec::discretization}

From Equations \eqref{eq:v} and \eqref{eq:measurement}, we can decompose the observation into two sets of integrals:

\begin{align}
    y_i^{(o)} &= \left[\frac{1}{|D_i^{(o)}|}\int_{D_i^{(o)}}g_i(s)x(s)ds\right]^\top \beta + \frac{1}{|D_i^{(o)}|}\int_{D_i^{(o)}}g_i(s)r(s)ds+\varepsilon_i \notag\\
    &= x_i^{(o)\top}\beta+r^{(o)},\label{eq:two_terms}
\end{align}

with $x_i^{(o)} = \frac{1}{|D_i^{(o)}|}\int_{D_i^{(o)}}g_i(s)x(s)ds$ and $r_i^{(o)} = \frac{1}{|D_i^{(o)}|}\int_{D_i^{(o)}}g_i(s)r(s)ds.$

It is convenient to handle the covariate and residual terms separately when approximating these integrals.  
First, the covariates are observed directly, whereas the residual field \(r\) is unobserved and must be inferred; simple averaging or quadrature or even  measurements can be used to obtain \(x^{(o)}\), but no such shortcut exists for \(r^{(o)}\), which requires estimation of the latent process.  
Second, covariates are often available at much finer resolution than the latent field. Allowing distinct discretizations lets us exploit this higher spatial resolution for \(x\) without incurring the computational cost of a fine discretization for \(r\).  
Finally, theoretical analyses of discretization error for random fields typically assume a constant mean, an assumption violated once covariates---often not even continuous functions---are included.  
For these reasons, we discretize \(x\) and \(r\) separately.
 
\subsubsection{Discretization of the residual}\label{sec:disc_resid}

The more challenging term in Equation \eqref{eq:two_terms} to approximate is $r_i^{(o)}$. With a mutually exclusive and exhaustive latent tiling $\{D_j^{(\ell)}\}_{j=1}^m$ of $D$, each observation decomposes into integrals over intersections $D_i^{(o)} \cap D_j^{(\ell)}$:

 \begin{align}
    r^{(o)} = \frac{1}{|D_i^{(o)}|}\sum_{j=1}^{n} \int_{D_i^{(o)} \cap D_j^{(\ell)}}g_i(s)r(s)ds.\notag
 \end{align} 
 By approximating these integrals, we can relate $r_i^{(o)}$ to $r_j^{(\ell)}$ linearly. Let $g_{ij}$ be the average of $g_i$ on $D_i^{(o)} \cap D_j^{(\ell)}$. The approximation given by Equation \eqref{eq::approximation}
\begin{align}
\int_{D_i^{(o)} \cap D_j^{(\ell)}} g_i(s) \, r(s) \, ds \;\approx\;
g_{ij} \, \frac{|D_i^{(o)} \cap D_j^{(\ell)}|}{|D_j^{(\ell)}|} \int_{D_j^{(\ell)}} r(s) \, ds \label{eq::approximation}
\end{align}
is consistency-preserving  (i.e., the error vanishes as resolution increases). Using this, we can acquire the linear relation

\begin{align}
r^{(o)} \approx \sum_{j} a_{ij} \, r_j^{(\ell)}+\varepsilon_i, \quad
a_{ij} = \frac{|D_i^{(o)} \cap D_j^{(\ell)}|}{|D_i^{(o)}|} \, g_{ij}. \label{eq:discretized}
\end{align}
The approximation error is controlled by the field's roughness and the intersection geometry: a mean--squared error bound can be expressed in terms of the variogram and region sizes (as is proved in the Supplement, Section~1). Thus $A:= A_{ij} =  a_{ij}$ is a consistency-preserving discretization of the continuous change-of-support map, independent of any specific covariance family.

\subsubsection{Integrating the covariates} 

Equation \eqref{eq:discretized} shows that the likelihood for the coarse observations $y^{(o)}$ depends only on the covariates aggregated to the observation footprint. In particular, posterior inference for $\beta$ and $r^{(\ell)}$ does not require knowledge of the fine--resolution covariates $x^{(\ell)}$; those enter only when forming $y^{(\ell)} = X^{(\ell)}\beta + r^{(\ell)}$ as a linear transformation of the posterior in Equation \eqref{eq:posterior_yl}.

In practice, covariates are often available at a finer spatial resolution than the latent field. Unlike downscaling, aggregating a high-resolution covariate to a coarser scale is well-posed, and a standard approach is weighted averaging over the footprint (with weights given by the sensor response $g$ when known). For example, a 10 m land-cover mosaic can be averaged to a 330 m latent grid or to kilometer-scale footprints to produce $X^{(\ell)}$ and $X^{(o)}$. When covariates are defined at the latent resolution, the same change-of-support weights $A$ in \eqref{eq:discretized} provide a consistency-preserving approximation to $X^{(o)}$ under mild continuity conditions, paralleling the argument used for $r^{(\ell)}$.

\subsection{Bayesian Linear Model}\label{sec::blm}

Using the discretization in \eqref{eq:discretized} we construct a Bayesian linear model as follows:

\begin{align}
    y^{(o)} | X^{(o)},r^{(\ell)},\beta,\sigma^2 &\sim N(X^{(o)}\beta+Ar^{(\ell)},\sigma^2 \Sigma_\varepsilon),\label{eq:likelihood_yo}\\
    r^{(\ell)}|\sigma^2 &\sim N(0,\sigma^2(\tau Q)^{-1})\label{eq:prior_yl},\\
    \beta &\sim N(0,\Lambda_\beta^{-1}),\notag \\
    \sigma^2 &\sim \text{inv-gamma}(a,b) \notag.
\end{align}

In our SIF application, $\Sigma_\varepsilon$ is a diagonal matrix with entries given by the per-sounding uncertainty estimates reported by OCO-2, and $\Sigma_\varepsilon = I$ in the semi-synthetic validation where noise is homoskedastic by construction.

It is helpful to write this in block form, which condenses into a standard Bayesian linear regression form. Setting $\Phi = \begin{bmatrix} X^{(o)} & A \end{bmatrix}$, $\Lambda_{\beta,r} = \text{blockdiag}(\Lambda_{\beta},\tau Q)$, and $\gamma = \begin{bmatrix} \beta & r^{(\ell)}\end{bmatrix}$, the likelihood is

\begin{align}
    y^{(o)} | \Phi,\gamma,\sigma^2 &\sim N(\Phi\gamma,\sigma^2 \Sigma_\varepsilon), \notag\\
    \gamma | \sigma^2 &\sim N(0,\Lambda_{\beta,r}), \notag
\end{align}

with $\sigma^2$ unchanged. 

Following the standard Bayesian linear regression formulation, the posterior of $\gamma |\sigma^2,y^{(o)}$ is given by

\begin{align}
    \gamma | \sigma^2,y^{(o)} &\sim N(\mu_{\gamma},\Lambda_{\gamma}^{-1}) \notag
\end{align}

with

\begin{align*}
    \Lambda_{\gamma} &= \Phi^\top \Sigma_\varepsilon^{-1} \Phi+\Lambda_{\beta,r},\\
    \mu_{\gamma} &= \Lambda_{\gamma}^{-1}\Phi^\top \Sigma_\varepsilon^{-1} y^{(o)}.
\end{align*}

Inference on $y^{(\ell)} = X^{(\ell)}\beta+r^{(\ell)}$ then proceeds as a linear transformation of $\gamma:$ $y^{(\ell)} = \begin{bmatrix} X^{(\ell)} & I\end{bmatrix} \gamma.$ The posterior distribution of $y^{(\ell)}$ given $\sigma^2$ then is

\begin{align}
    y^{(\ell)} | \sigma^2,y^{(o)} \sim N\left( \begin{bmatrix} X^{(\ell)} & I\end{bmatrix}\mu_{\gamma},  \begin{bmatrix} X^{(\ell)} & I\end{bmatrix}\Lambda_{\gamma}^{-1} \begin{bmatrix} X^{(\ell)} & I\end{bmatrix}^\top\right). \label{eq:posterior_yl}
\end{align}

An advantage of this Bayesian framework is to resolve the unidentifiability that occurs when there are more latent grid cells than observations: without the prior, there are many $\hat{\gamma}$ which satisfy $\gamma = \arg \min_{\gamma} ||y^{(o)}-\Phi\gamma||$, and so a Bayesian approach resolves the question of how to choose an estimate $\hat{\gamma}$ so that $\hat{y}^{(\ell)}$ realistically behaves like a spatially correlated field. Using $\hat{y}^{\ell}$ as the posterior mean of the model in Equation \eqref{eq:likelihood_yo}, we can view this point estimate as solving the penalized regression problem
\begin{align}
    \hat{\gamma}=\arg\min_{\gamma} ||y^{(o)}-\Phi \gamma||_{\Sigma_\varepsilon^{-1}}+\gamma^\top\Lambda_{\beta,r}\gamma,\label{eq:penalty}
\end{align}
where $||\cdot||_{\Sigma_\varepsilon^{-1}}$ denotes the $\Sigma_\varepsilon^{-1}$-weighted norm.

The Bayesian formulation is not strictly necessary for the point estimate, which coincides with penalized least squares under plug-in $\sigma^2$. Its advantages are interpretive and extensible: the SAR prior provides a natural probabilistic motivation for the smoothness penalty; the posterior distribution provides a principled basis for uncertainty quantification; and full inference over $\sigma^2$ is available to users requiring complete uncertainty propagation, though the plug-in approximation is well-justified at the observation densities considered here.

The connection between \eqref{eq:penalty} and the model given by \eqref{eq:likelihood_yo} and \eqref{eq:prior_yl} demonstrates that the behavior of point estimates depends on the choice of $Q$: if $Q = I$ is chosen, giving prior precision $\tau I$, this corresponds to a prior belief that there is no spatial correlation between $r^{(\ell)}$, and resulting maps have grainy patterns (the differences between adjacent pixels can be large), while spatially correlated $Q$ prefer smoother fields. In Section \ref{sec::prior} we propose a spatially correlated $Q$ based on SAR models, and in Section \ref{sec::hyperparameter} we discuss how we use an empirical-Bayes approach to tune $\tau$ using cross-validation.

\subsection{Prior Specification}\label{sec::prior}

To appropriately incorporate spatial structure into the prior on $r^{(\ell)}$, we use a simultaneous-autoregressive (SAR) precision structure. This structure arises when each cell's value is modeled as linear combination (typically an average) of its neighbors:

\begin{align}
    r^{(\ell)} = Wr^{(\ell)}+\varepsilon^{(\ell)}, \notag
\end{align}

giving the structure matrix

\begin{align}
    Q = (I-W)^{\top}(I-W), \label{eq::prior_specification}
\end{align}
where $W$ denotes an adjacency matrix, so that the prior precision for $r^{(\ell)}$ is $\tau Q$. While there are many ways to define an adjacency matrix, in our analyses we focus on a common form. First, let $\tilde{W}$ denote the matrix where $\tilde{W}_{ij} = 1/n_i$ if pixel $i$ and pixel $j$ are neighbors and 0 otherwise and where $n_i$ is the number of neighbors of pixel $i$. We use the ``queen" adjacency (so that interior pixels have 8 neighbors). It is common to set $W = \rho \tilde{W}$ for $\rho \in (0,1]$. When $\rho = 1$, $Q$ is not full-rank and $\tau Q$ is improper (resulting in an intrinsic SAR prior). Since $W$ is row-stochastic and its associated adjacency graph is connected, $\mathbf{1}$ is the unique unit eigenvalue of $W$, and $I - W$ has rank $m - 1$. The posterior is nonetheless proper provided every pixel has at least one neighbour, as the posterior precision borrows rank from $A^\top A$; a proof is given in the Supplement (Section~1). An intrinsic prior is appealing for reducing model complexity; in our semi-synthetic experiments in Section \ref{sec::SyntheticAlbedo}, setting $\rho = 1$ produced nearly identical predictions and uncertainty estimates compared to other plausible values of $\rho$ (values of $\rho$ with good cross-validation performance, see Section \ref{sec::hyperparameter}), but nevertheless we report results after tuning $\rho$ in this paper. 

Combined with covariates, this formulation induces a prior on $y^{(\ell)} |\beta$ equivalent to a common formulation for SAR models. Here,

\begin{align}
    r^{(\ell)} &= y^{(\ell)}-X^{(\ell)}\beta \notag
\end{align}

is modeled using the SAR formulation, which is a common specification  \citep{hooten2014simultaneous}. Thus, this model is consistent with a conventional SAR prior with covariates on $y^{(\ell)}$.

In our experiments, we also consider a landcover-aware version of $W$ that reduces communication between land and water pixels. However, this did not have a substantial impact on the downscaling results after the landcover was included as a covariate in the model; full results are provided in the Supplement (Section~3).

As for the prior for $\sigma^2$, we use a non-informative Jeffrey's prior (setting $a = b = 0).$

\subsection{Posterior Inference}
Posterior inference in this model without covariates is otherwise routine: given the Gaussian
likelihood and Gaussian SAR prior, full conditional
distributions are available in closed form.
Because the observational variance $\sigma^2$ is tightly identified with many
observations, we condition on its posterior mode $\hat\sigma^2$ rather than
sampling. With a larger number of observations, this plug--in step yields practically identical posterior summaries
to full Bayesian inference, while greatly reducing computational cost. 

\subsubsection{Identifiability and Spatial Confounding}\label{sec::confounding}

Including covariates introduces a familiar identifiability problem: should the
model explain variability through $x(s)^\top \beta$ or through the residual
field $r(s)$? This spatial confounding \citep{Hodges01112010} can obscure
inference on $\beta$ and, by extension, on $y^{(\ell)}$.

In our application, the land--water boundary creates a sharp spatial contrast
that both the water proportion covariate and the SAR random effect can absorb.
Without any correction, the estimated water coefficient is positive --- implying
water pixels have higher SIF than vegetated land --- which is physically
implausible and inconsistent with the negative relationship observed at the
coarse sounding scale ($\hat{\beta}_{\text{water}} \approx -0.67$, Supplement Section~4). \citeauthor{Hodges01112010} suggest restricted spatial regression
(RSR), which orthogonalizes the residual field against the covariate, yielding
coefficients $\beta$ identical in distribution to fitting the model without the
spatial random effect. We adopt RSR as a pragmatic design choice for this
application rather than as a general recommendation. Enforcing the constraint
$X^{(o)\top} A r^{(\ell)} = 0$ via the Schur complement of the constrained
Gaussian amounts to two additional linear solves and a $2 \times 2$ matrix
inversion when there is a single covariate --- negligible overhead relative to
the posterior mean computation.

\subsection{Hyperparameter Selection}\label{sec::hyperparameter}

We selected the smoothing parameter $\tau$ by $K$-fold cross-validation (with $K = 10)$, minimizing RMSE on held-out folds. Soundings were assigned to folds with a simple random sample with no spatial blocking. This procedure can be interpreted as an empirical Bayes approach, providing a plug-in estimate of $\tau$ that balances bias and variance. 

\subsection{Boundary Effects and Buffering}

Because the SAR prior is defined through local neighborhoods, boundary pixels have fewer neighbors than interior pixels. This leads to weaker prior penalization at the edges of the domain and consequently larger posterior variance near boundaries---a phenomenon sometimes referred to as a ``boundary effect.'' To mitigate this, we embed the target grid using a buffer of 25 pixels. 25 is chosen experimentally as the smallest number of pixels where adding an additional 5-pixel buffer changes all predictions in the target area by less than 1\%. 

When using an intrinsic SAR prior, there is an implicit flat prior on the intercept. This induces global correlation through the intercept term, which limits the usefulness of general guidelines for choosing a buffer size. Empirical checking is therefore a practical approach.

\section{Semi-Synthetic Albedo Validation}\label{sec::SyntheticAlbedo}
To evaluate model performance on a known dataset, we generate semi-synthetic data by upscaling a 10 m albedo product to match the OCO-2 SIF sounding geometries centered over Boston. This setup allows us to assess the method by quantifying how accurately it recovers the original fine-resolution grid used to generate the data. Boston is chosen as an example case for all of this study as it has heterogeneous urban landcover and a high density of target-mode observations (see Figure \ref{fig:downscaling_plots_ndvi_se}).

\begin{figure}
\centering
\begin{tikzpicture}[remember picture]
  \node (A) {\includegraphics[width=0.33\linewidth]{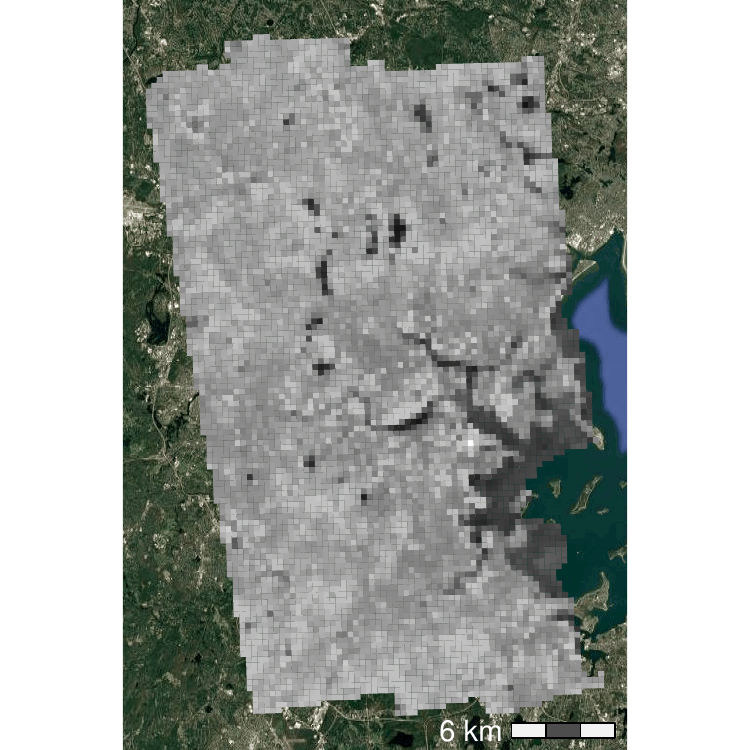}};
  \node[right=-10mm of A] (B) {\includegraphics[width=0.33\linewidth]{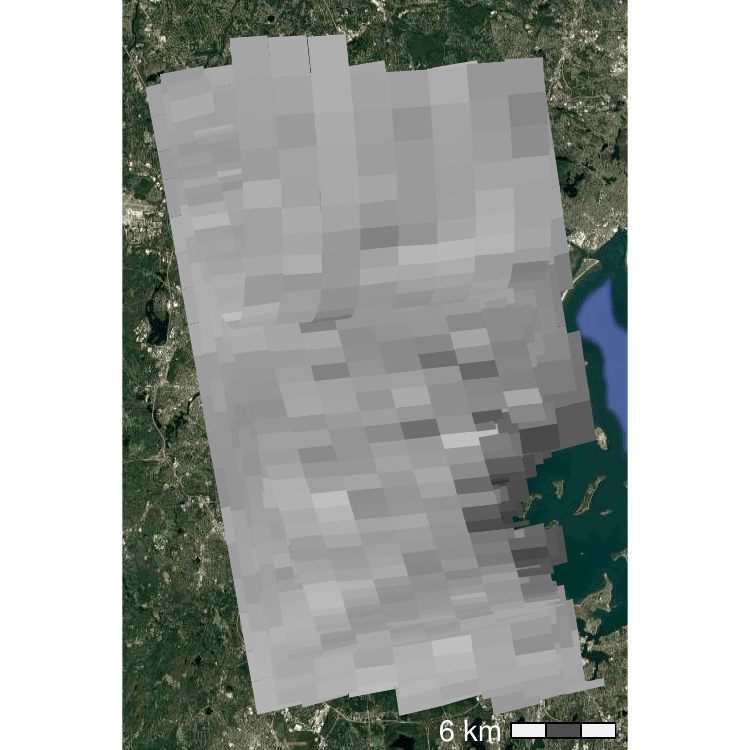}};
  \node[right=-10mm of B] (C) {\includegraphics[width=0.33\linewidth]{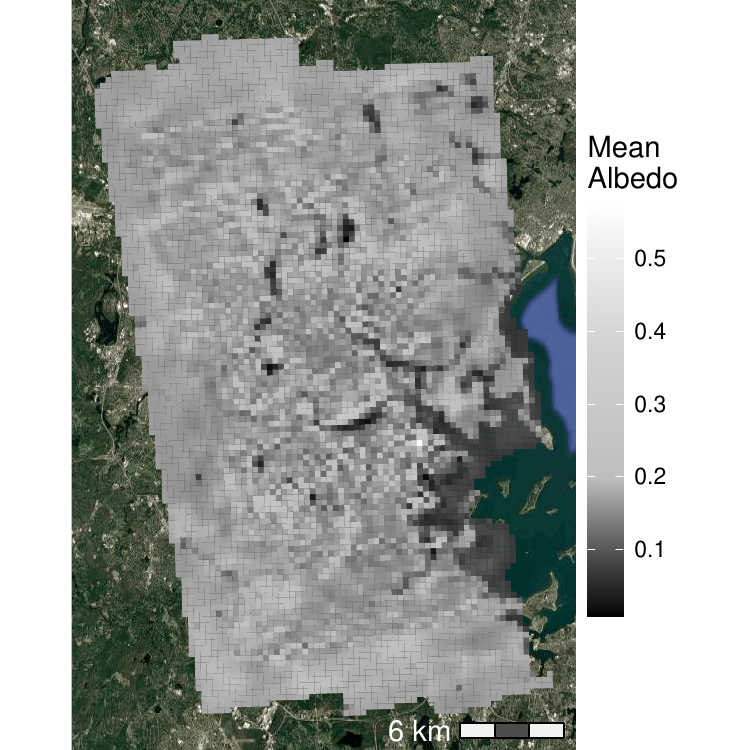}};
  \coordinate (Cright) at (C.east);
\end{tikzpicture}
\caption{Semi-synthetic albedo downscaling results. Derived mean albedo product (left); upscaled semi-synthetic albedo measurements (middle), generated by averaging 10 meter Sentinel-2 albedo over each OCO-2 sounding footprint and adding noise; posterior mean of the downscaled field (right).}
\label{fig:combined_downscaling_plots}
\end{figure}

\begin{figure}
    \centering
    \includegraphics[width=1\linewidth]{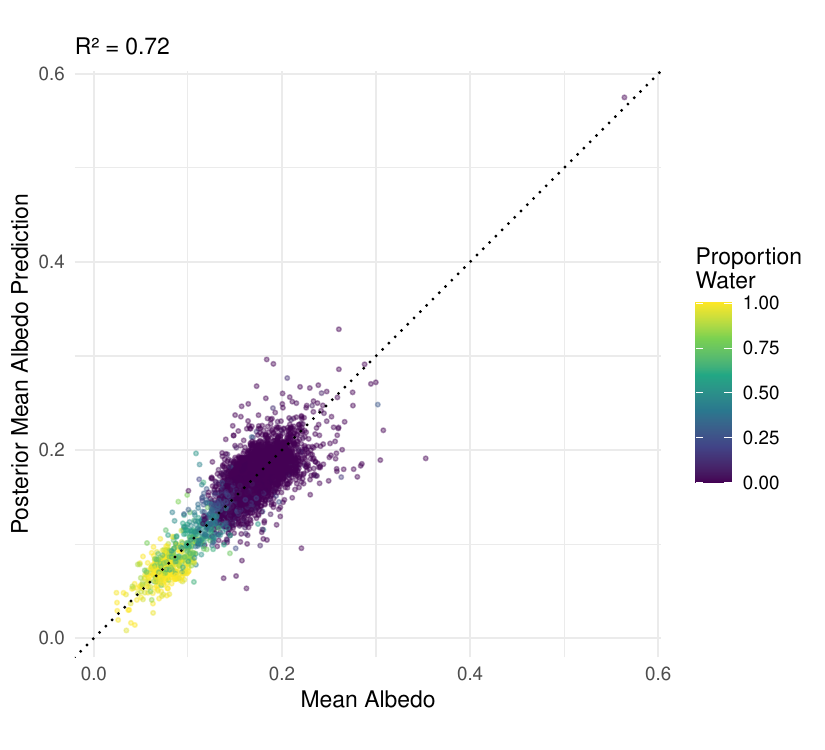}
    \caption{True albedo product are plotted against posterior mean predictions, colored by proportion water plotted along with the line $y = x$ (dotted line): the predictions linearly relate to the generating field, and water is cleanly separated in the predictions with mixed water-fraction pixels increasing with a gradient. The greatest outlying points correspond to isolated buildings or pixels touching a convention center. This relationship has an $R^2$ of 0.72 (see Table \ref{tab:ablation_table}).}
    \label{fig:truth_vs_predictions_albedo}
\end{figure}

To generate the synthetic dataset, we begin with albedo derived from Sentinel-2 using the algorithm by \cite{Bonafoni} for a June 2022 cloud-free median composite taken over Boston at the 10 meter resolution. Next, these data are aggregated from 10 meter to the coarse geometries using the mean value, which corresponds to a sensitivity distribution of $g_i(s) = 1$. We added i.i.d normal noise to simulate measurement error, setting the standard deviation of $\varepsilon_i$ to 5\% of the standard deviation of the albedo product; accordingly, $\Sigma_\varepsilon = I$ throughout this validation. Because the synthetic observations are generated by direct aggregation of 10 meter albedo over each sounding footprint, the data-generating integral is resolved at finer resolution than the 330 meter aggregation matrix
$A$ approximates. The validation therefore assesses recovery against a more accurate integral than the one used in inference — a stricter test than the correctly-specified case, in which $A$ would exactly reproduce the data-generating process.

To obtain a dataset for landcover, we aggregate landcover classes from the ESA WorldCover 2020 product \citep{zanaga2022esa} at the 10 meter resolution by assigning each raster pixel to the dominant (mode) landcover class. We combine the classes from the ESA WorldCover product into a harmonized land cover nomenclature using the scheme in Table \ref{tab:landcover}. This dataset is used for validation as well as to generate a covariate: $x(s) = 1$ if the class is water, and 0 otherwise. With this covariate, $\int_{D} x(s)ds$ is then the proportion of a region that taken up by water, which can be computed by averaging the number of 10m pixels are water in the region.

\begin{table}[ht]
\centering
\caption{Predictive performance for the semi-synthetic albedo ablation scheme. 
RMSE and $R^2$ are computed against the true latent field; coverage is 
approximate 95\% credibility interval coverage with plug-in mode for $\sigma^2$. 
The coarse-scale OLS baseline applies ordinary least squares to the aggregated 
data and predicts directly to the grid without spatial structure; no coverage 
is reported as it produces point predictions with no principled uncertainty 
quantification. Universal kriging treats observations as point support at 
sounding centroids, ignoring the areal nature of the footprints.$^\dagger$}
\label{tab:cv_summary}
\label{tab:ablation_table}
\begin{tabular}{lcccc}
\toprule
Model &  & RMSE & $R^2$ & Coverage (95\%) \\
\midrule
No covariates &  & 0.0208 & 0.5692 & 0.9702 \\
Include Water &  & 0.0170 & 0.7115 & 0.9720 \\
Include Water (orthogonalized resid) &  & 0.0168 & 0.7180 & 0.9736 \\
Coarse-scale OLS baseline &  & 0.0201 & 0.5978 & \textemdash{} \\
\midrule
Universal kriging (point support)$^\dagger$ &  & 0.0165 & 0.729 & 0.780 \\
\bottomrule
\end{tabular}

\smallskip
\raggedright\small $^\dagger$ Kriging predictions are made at 330 m grid cell centroids using an empirically fitted exponential variogram; coverage reflects interpolation uncertainty only and does not account for change-of-support uncertainty. The supplement (Section 6.3) reports a corresponding kriging coverage of 72.4\% under self-consistent generation; the difference from the 78.0\% reported here reflects the stricter 10 m generation procedure described in Section 3.
\end{table}

\begin{table}[h] 
    \centering
    \begin{tabular}{lll}
        \toprule
        \textbf{Landcover Class} & \textbf{ESA WorldCover Label} & \textbf{  Study Region Proportion} \\
        \midrule
          Forest & Tree cover, Mangroves  & 53.9 \%\\
        Urban & Built up & 27.7\%\\        Water & Permanent water bodies & 8.6\% \\
        Grass & Grassland, Moss and lichen  & 6.0 \%\\  
        Bare Surface & Bare / sparse vegetation & 2.3 \% \\ 
        Wetlands & Herbaceous wetland & 1.4\% \\           
    \end{tabular}
    \caption{Landcover Class Labels and corresponding ESA WorldCover Labels. Other labels, including croplands, ice and snow, and shrub and scrub, had less than 0.1\% representation in the study region.}
    \label{tab:landcover}
\end{table}

To assess how different components of the model contribute to predictive performance, we constructed an ablation scheme that incrementally adds structure. As a baseline, we include a coarse-resolution regression approach: this procedure is the simplest regression procedure described by \cite{park2019overview} in which the covariate is upscaled to the observation geometries (through simple averages or aggregates over the observation footprint), an ordinary least squares regression is fit without any spatial structure, and then predictions are made on the 330 meter resolution. This model serves as a baseline to compare our approach to an overall intercept and one major structural component. We then  downscale with the simultaneous autoregressive prior described in Section \ref{sec::prior}, tuning both $\rho$ and $\tau$ as hyperpareters with cross-validation. While we do try other values of $\rho$, we find that the difference between the intrinsic $\rho = 1$ and other competitive values of $\rho$ are minor and so we leave those in the Supplement (Section~7). Next, we augment the downscaling model with proportion of water as a covariate to evaluate the role of auxiliary predictors as in Section \ref{sec::blm}. We compare this result to the restricted version from Section \ref{sec::confounding}, restricting the spatial noise component to be orthogonal to the covariate component to improve identifiability. For each model variant we report results at the cross-validated minimum smoothing parameter; in all experiments we find that this yields near-nominal or mildly conservative coverage (95--98\% for marginal, pixel-level 95\% credible intervals), with the slight overcoverage reflecting the model's conservative absorption of forward operator approximation error into $\sigma^2$, and so we leave perturbations to the Supplement (Section~6). In general, lower smoothing parameter means increased coverage. 

Each model was assessed by three metrics. The posterior mean of the downscaled field was used as a prediction at the 330 meter scale where data-generating means are known. At this scale, we assess the RMSE, and we assess the $R^2$ value obtained when regressing the generating field against the prediction in a simple linear model with intercept. We also assess approximate coverage in the following way: for each pixel, an approximate 95\% credibility interval is constructed using its posterior mean and variance, that is, (given by $\hat{y}^{(\ell)} \pm 1.96\sqrt{\hat{\sigma}^2\Lambda^{-1}_{ii}}$, with $\hat{\sigma}^2$ the mode of the posterior distribution for $\sigma^2$ and $\Lambda$ the posterior precision matrix). Then, ``approximate coverage" is defined as the proportion of pixels for which the true value is within that 95\% interval. Results from this scheme can be seen in  Table \ref{tab:ablation_table}.

Predictive performance is broadly similar across model variants in the albedo
experiment --- the full model achieves RMSE of 0.0168 compared to 0.0208
without covariates and 0.0201 for the OLS baseline --- suggesting that for
albedo, RSR's primary benefit is coefficient interpretability rather than
predictive accuracy. Universal kriging at point support achieves slightly better
RMSE (0.0165) but severely undercovering credible intervals (78\% vs.\ nominal
95\%), illustrating that ignoring the areal nature of the observations primarily
damages uncertainty quantification rather than point prediction --- a pattern
consistent with the misspecification results in the Supplement (Section~6). In the SIF
application, however, the spatial relationship between vegetation and water
boundaries is more complex, as green vegetation tends to occur near water,
making footprint-averaged observations particularly susceptible to confounding;
RSR is therefore expected to provide greater predictive benefit in that setting.

These simulation results inform our choices in the application to SIF in section \ref{sec::sif_application}.

\begin{figure}
    \centering
    \begin{minipage}[t]{0.32\linewidth}\vspace{0pt}
  \includegraphics[width=\linewidth]{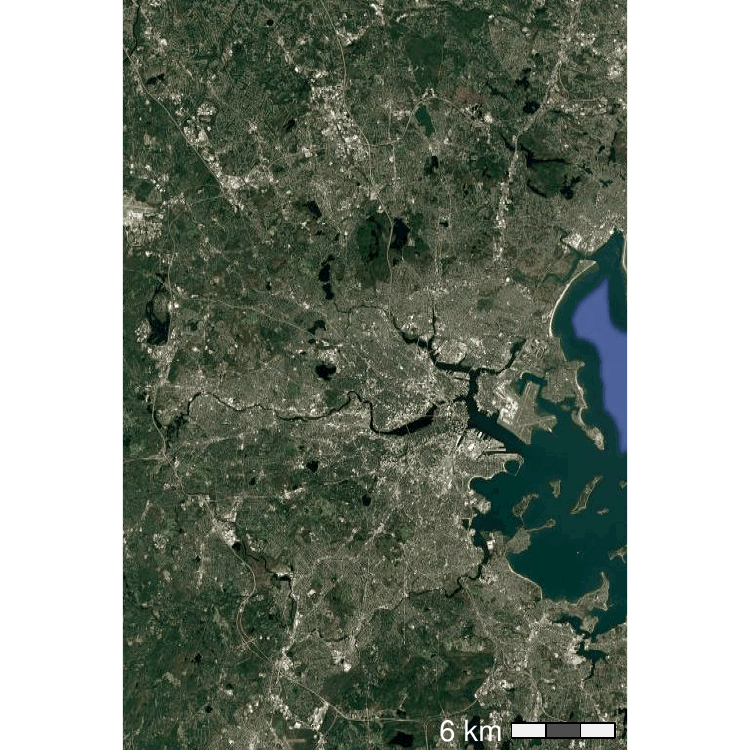}
\end{minipage}\hfill
\begin{minipage}[t]{0.32\linewidth}\vspace{0pt}
  \includegraphics[width=\linewidth]{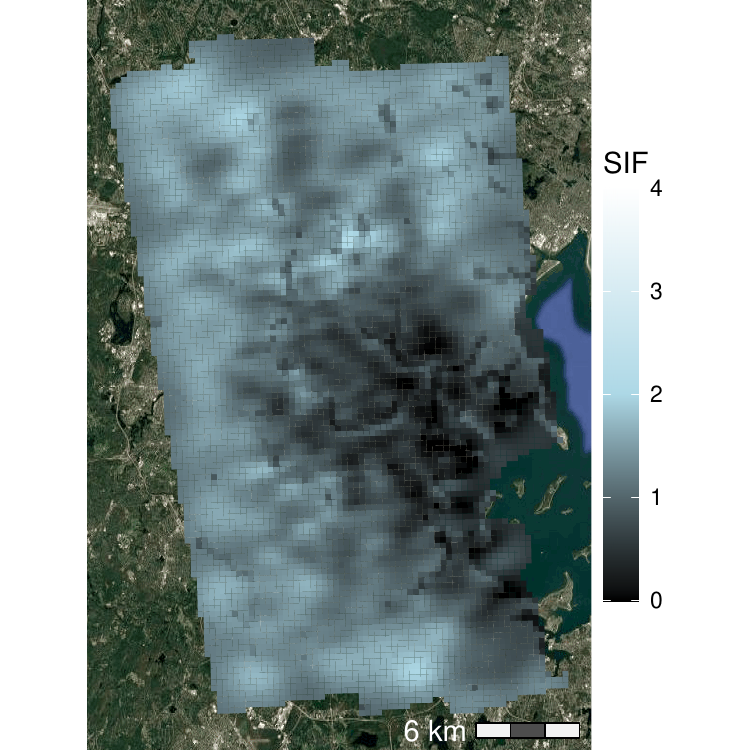}
\end{minipage}\hfill
\begin{minipage}[t]{0.32\linewidth}\vspace{0pt}
  \includegraphics[width=\linewidth]{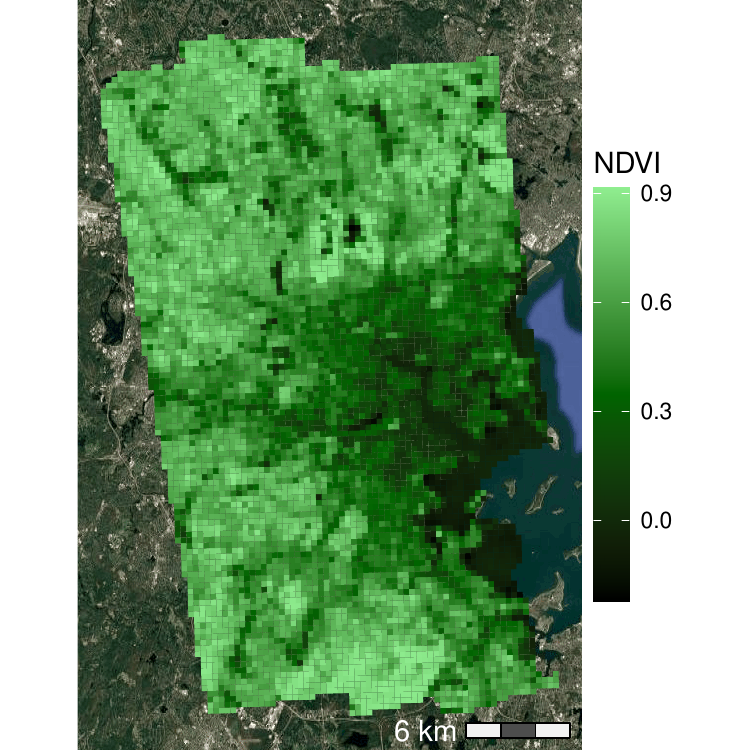}
\end{minipage}\hfill

\vspace{0.6em}

\begin{minipage}[t]{0.32\linewidth}\vspace{0pt}
  \includegraphics[width=\linewidth]{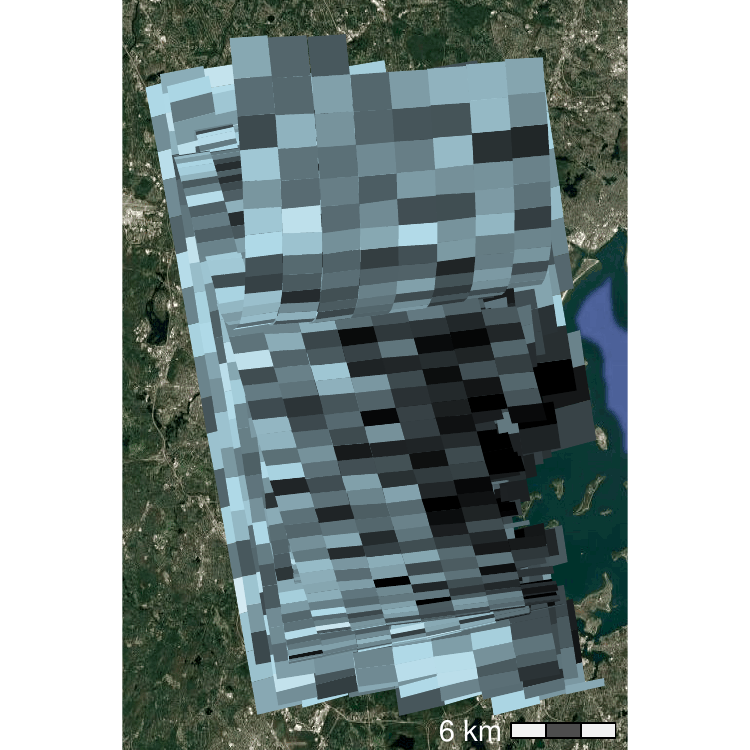}
\end{minipage}\hfill
\begin{minipage}[t]{0.32\linewidth}\vspace{0pt}
\hspace{-0.27em}\includegraphics[width=\linewidth]{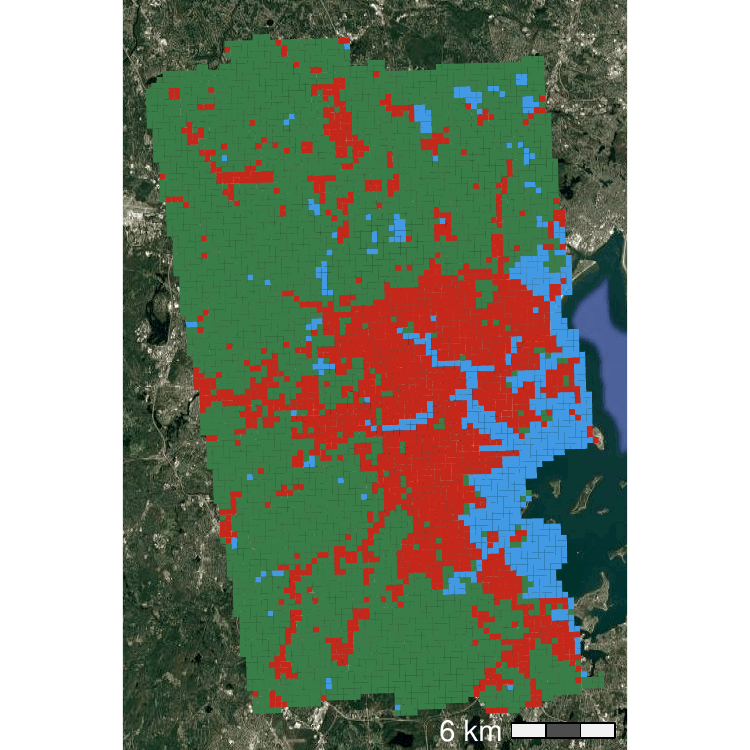}
\end{minipage}\hfill
\hspace{0.02em}\begin{minipage}[t]{0.32\linewidth}\vspace{0pt}
  \includegraphics[width=\linewidth]{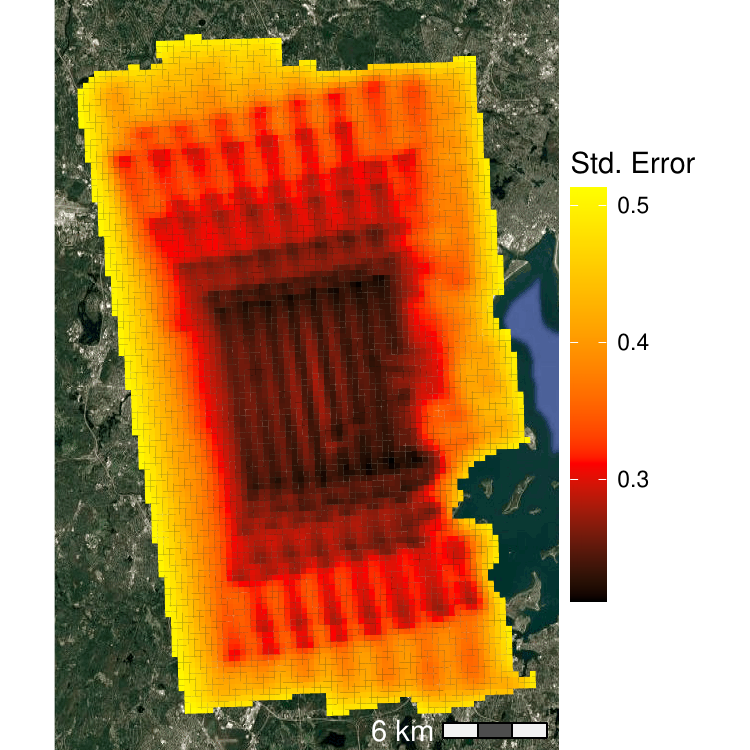}
\end{minipage}

    \caption{Boston satellite view (top left); posterior mean after downscaling (top middle); mean NDVI (top right); sounding SIF observations (SIF$_{757}$ [mW m$^{-2}$ nm$^{-1}$ sr$^{-1}$]); landcover (bottom middle with water or wetlands in blue, urban or bare surface in red, and forest, grass, shrub \& scrub or croplands in green); posterior standard error of downscaled SIF (right). }
    \label{fig:downscaling_plots_ndvi_se}
\end{figure}

\section{Application to OCO-2 SIF Data}\label{sec::sif_application}

We downscale OCO-2 SIF across Boston \citep{oco2siflite} and assess the degree to which the product reflects vegetation patterns. The target grid and sounding geometries are the same as in Section \ref{sec::SyntheticAlbedo}. We use water as a covariate, use the SAR prior with ``queen" adjacency weights discussed in Section \ref{sec::prior}, apply the restricted-regression orthogonalization approach from Section \ref{sec::confounding}, and select the scale parameter using 10-fold cross-validation. Per-sounding OCO-2 uncertainty estimates are used to construct $\Sigma_\varepsilon$ as described in Section~\ref{sec::blm} \citep{oco2siflite}.

We use two external datasets to validate the downscaling results. First, to compare the downscaling results with a measure of greenness, we average a 10 meter resolution Normalized Difference Vegetation Index (NDVI)  product to the 330 meter target. To compare results by landcover type, we use land-cover proportions from the ESA WorldCover product (Table 
\ref{tab:landcover}). 

\begin{figure}[H]
    \centering
    \includegraphics[width=0.49\linewidth]{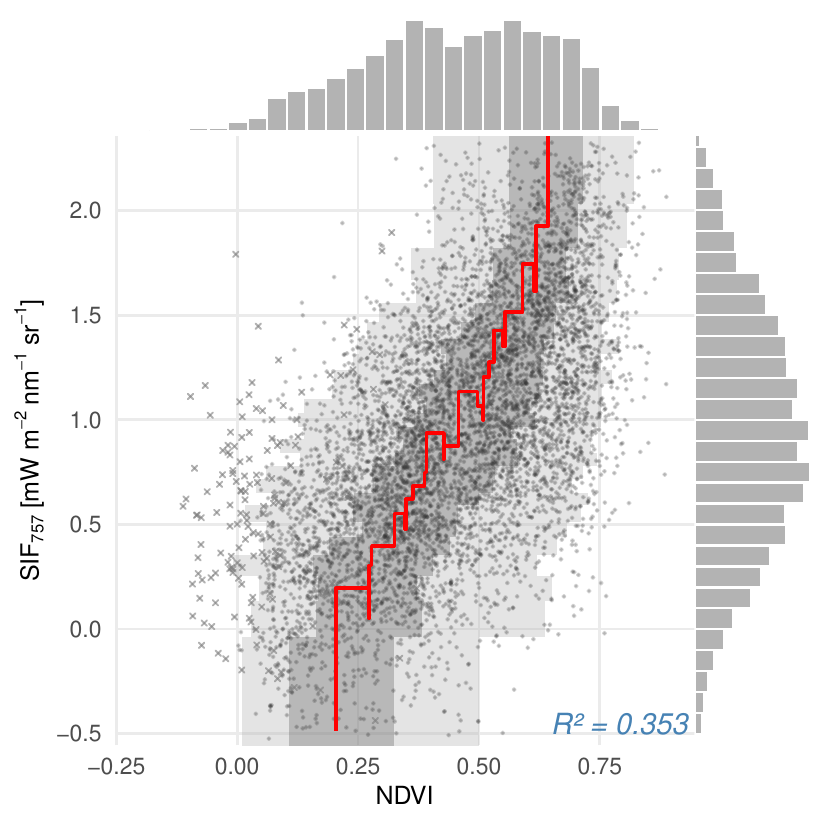} 
    \includegraphics[width=0.49\linewidth]{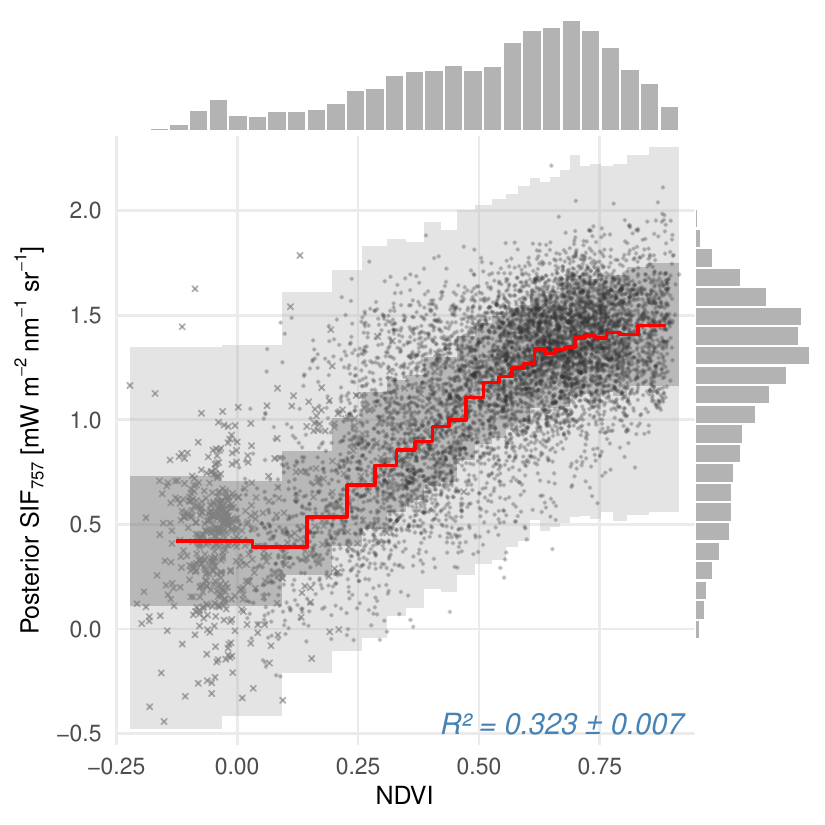}
    \caption{Relationship between Solar-Induced Fluorescence (SIF$_{757}$[mWm$^{-2}$nm$^{-1}$sr$^{-1}$]) and NDVI at (a) the coarse-resolution sounding scale and (b) the fine-resolution posterior grid scale. Points show observed (a) or posterior-mean (b) SIF; pixels with $>$ 50 \% water coverage are plotted as crosses. Shaded bands represent quantiles of the conditional distribution (0.025--0.975 light gray; 0.25--0.75 dark gray), and red lines mark the bin medians. In (a), quantiles summarize empirical NDVI variation within SIF bins. In (b), quantiles summarize posterior SIF variation within NDVI bins. Posterior $R^2$ values are computed for many posterior draws of SIF $\sim$ NDVI; the mean $\pm$ two standard deviations are shown to indicate uncertainty in the overall explanatory strength.}
    \label{fig:sif_ndvi_plot}
\end{figure}

The relationship between the posterior distribution of SIF and NDVI is shown in Figure~\ref{fig:sif_ndvi_plot}.  
To characterize the structural relationship between SIF and NDVI, we use quantiles and samples from the posterior distribution. The quantiles suggest an overall positive  correspondence between the downscaled SIF and greenness. Through posterior sampling (10,000 samples decorrelated by taking only the diagonal of the posterior covariance), we obtain an $R^2$ of $0.323 \pm 0.007$ between the downscaled SIF and NDVI. It is important to note that many of the pixels correspond to only a small number of overlapping soundings, so the strength of this relationship improves when considering pixels with more overlap, as posterior uncertainty decreases with observation density. For instance, if we consider only pixels that have at least 20 overlapping soundings, the $R^2$ value is $0.417 \pm 0.012$.

In the maps in 
Figure \ref{fig:downscaling_plots_ndvi_se}, we see relative alignment between SIF and NDVI, and a heatmap of posterior standard errors shows alignment of errors with observation density. 

In this analysis, we chose to limit the covariates to only one (water proportion) in order to validate the downscaling methodology; the model supports the addition of more covariates if prediction accuracy is prioritized.

\section{Discussion}

\subsection{Interpretation of uncertainty estimates}

The availability of calibrated uncertainty estimates distinguishes 
this framework from geometry-informed oversampling approaches 
such as \citet{turner2020double}. Posterior uncertainty directly 
reflects the degree to which the overlapping sounding geometry 
constrains each pixel's mean value: pixels with many overlapping 
soundings have tight posterior distributions, while pixels with 
few have wide ones, with the SAR prior filling in by borrowing 
strength from neighbors, as illustrated in Figure~4. The framework 
provides a complete Bayesian specification in which $\sigma^2$ can 
be treated as a parameter rather than fixed at its posterior mode; 
users requiring full uncertainty propagation over $\sigma^2$ can 
do so at modest additional computational cost.

It is worth being explicit about what the posterior intervals do and do not
capture. The inferential targets here are cell means rather than point values,
which means---unlike in point-support prediction where aleatoric uncertainty
sets an irreducible predictive floor---the posterior uncertainty is purely
epistemic and vanishes with observation coverage. Because we condition on the
posterior mode of $\hat{\sigma}^2$ rather than marginalizing over it, the
reported intervals reflect epistemic uncertainty in the latent field given a
fixed noise level, but do not propagate uncertainty in $\sigma^2$ itself.
Under a Jeffreys prior the posterior of $\sigma^2$ is inverse-gamma with shape
$n/2 \approx 3{,}500$ and scale $(n/2)\hat{\sigma}^2$; in our application
$\hat{\sigma}^2 \approx 0.17$ with a posterior standard deviation of
approximately $0.003$, justifying the plug-in approximation in this setting.
In settings with fewer observations, full posterior inference over $\sigma^2$
would be preferable. The same plug-in logic applies to $\tau$, which is fixed
at its cross-validated value rather than marginalized over; uncertainty in
$\tau$ would widen the reported intervals, though the cross-validation surface
is relatively flat near its optimum. Finally, the intervals do not account for
misspecification of the aggregation operator $A$: departures from the assumed
uniform sensor response can induce mild undercoverage in regions of irregular
footprint geometry, as explored in the Supplement (Section~6).

\subsection{The SAR prior and the bias-variance tradeoff}

The simultaneous autoregressive prior plays the role of a spatial 
regularizer, encoding the belief that neighboring pixels should have 
similar values. The smoothing parameter $\tau$ controls the strength 
of this regularization: large $\tau$ enforces strong spatial smoothness 
at the cost of local detail, while small $\tau$ allows sharper 
pixel-to-pixel variation at the risk of overfitting to noise. We 
select $\tau$ by cross-validation, minimizing held-out RMSE on the 
observed soundings, which gives the procedure an empirical Bayes 
flavor analogous to selecting a penalty in ridge regression.

One potential concern with random cross-validation in a spatial 
setting is that nearby soundings share information about the latent 
field, allowing spatially proximate training points to inform 
held-out predictions and potentially underestimating generalization 
error. We investigated spatially blocked cross-validation as an 
alternative. In a separate diagnostic using the simpler, 
self-consistent generation procedure (Supplement, Section~2), 
spatially blocked cross-validation produces substantially 
undercovered credible intervals (82.2\%), while random cross-validation 
remains near-nominal (94.8\%). We therefore prefer random 
cross-validation for the main analysis; a detailed comparison of the 
two schemes, including the mechanisms that plausibly drive this 
discrepancy, is provided in the Supplement (Section~2).

As an alternative to cross-validation, $\tau$ can also be selected 
by maximizing the marginal likelihood, which is feasible at these 
scales and yielded similar results in our experiments. This provides 
reassurance that the cross-validation procedure is not overfitting 
to the particular fold structure, and that the selected $\tau$ 
reflects a genuine signal in the data rather than an artifact of 
the validation scheme.

\subsection{Scope and limitations of the semi-synthetic validation}

The semi-synthetic albedo experiment was designed to evaluate the statistical framework under controlled but geometrically realistic conditions. Synthetic soundings are generated by integrating 10 m Sentinel-2 albedo over the actual OCO-2 footprint geometries, so the features that make inference challenging — the degree of overlap, footprint shapes, the ratio of observations to grid cells, and genuine sub-grid variability — match those in the real application. The validation therefore speaks directly to the geometric and inferential machinery of the method.

These geometric strengths aside, albedo and SIF differ in their spatial structure, their relationship to land cover, and the processes governing sub-pixel variability; strong performance on albedo does not imply equivalent performance on SIF. The uniform sensitivity distribution $g_i(s) = 1$ is used throughout, whereas real OCO-2 soundings have a non-uniform response; sensitivity to this assumption is assessed in the Supplement (Section~6). The noise level — 5\% of the 330 meter albedo standard deviation — is likely smaller than true OCO-2 SIF measurement uncertainty. Counterintuitively, however, lower noise makes uncertainty quantification harder rather than easier, as measurement error provides smoothing that partially compensates for limited overlap; the near-nominal coverage obtained in this setting is therefore encouraging rather than a best-case result. The calibration results should be understood as demonstrating internal consistency of the inferential machinery -- that uncertainty accounting is correct, including the approximation error incurred by discretization -- rather than as a guarantee of performance under the full complexity of the SIF application.

\subsection{Dependence on sampling density and applicability 
beyond target mode}

Although this paper is motivated by the unusually dense target-mode 
acquisitions of OCO-2, the framework is not restricted to this 
setting. In sparser observation regimes the SAR prior contributes 
more to the posterior, predictions are smoother, and uncertainty 
intervals are wider---behavior that reflects the actual information 
content of the data rather than a limitation of the method. In this 
sense a probabilistic framework is arguably most valuable precisely 
when data are sparse, since it makes explicit what point-estimate 
approaches obscure: that fine-scale inference is poorly supported 
by the observations.

The main practical caveat for sparse settings is the plug-in 
$\sigma^2$ approach, which relies on $\sigma^2$ being tightly 
identified from the data. With few observations this assumption 
breaks down and we recommend retaining $\sigma^2$ as a parameter 
and integrating over it. Subject to this caveat, the method is 
broadly applicable to any satellite product with irregular areal 
footprints, including non-target-mode OCO-2 data, TROPOMI, or 
other instruments where overlapping soundings are available.

\subsection{Modelling assumptions and extensions}

Several simplifying assumptions merit acknowledgment. We assume 
homoskedastic measurement error in the semi-synthetic albedo 
experiment, where noise is injected at a known fixed level by 
construction, and accordingly set $\Sigma_\varepsilon = I$ in that setting. 
As noted in Section~\ref{sec::blm}, heteroskedastic per-sounding uncertainties 
are incorporated through $\Sigma_\varepsilon$ in the SIF application; 
we found modest impact on posterior means and standard errors relative to
the homoskedastic specification.

The neighborhood structure of the SAR prior is fixed to queen 
adjacency with equal weights. We also evaluated a landcover-aware 
variant that down-weights edges crossing land-water boundaries, 
finding negligible impact once water proportion was included as 
a covariate. Future work could explore richer neighborhood 
structures, including distance-weighted or anisotropic adjacency 
schemes, particularly for settings where the latent field is 
expected to exhibit directional structure.

Spatial confounding is addressed through restricted spatial 
regression, which orthogonalizes the residual field against the 
covariates. We note that RSR has been criticized for producing 
biased covariate estimates and potentially understating uncertainty 
when the true data-generating process does not respect the 
orthogonality constraint \citep{khan2022restricted, gilbert2025consistency}. 
In our setting this concern is secondary for two reasons: first, 
the primary inferential target is the downscaled field $y^{(\ell)}$ 
rather than $\beta$; and second, the sign reversal observed without 
RSR provides empirical evidence that the 
unconstrained model is itself misspecified in this application. 
The impact of RSR on posterior maps and coverage is documented in 
Table~\ref{tab:ablation_table}, where the orthogonalized model yields 
near-identical RMSE and marginally improved coverage relative to the 
unconstrained version. The framework is readily adaptable to other 
satellite-derived products with irregular areal footprints, such as 
column-averaged CO$_2$ (XCO$_2$) from OCO-2, and provides a foundation 
for future analyses requiring principled uncertainty quantification 
at fine spatial scales.

\section{Conclusion}

We have presented a Bayesian framework for downscaling satellite observations
with irregular, overlapping areal footprints, motivated by OCO-2 target-mode
SIF acquisitions over Boston. The central technical contribution is a sparse,
consistency-preserving discretization of the change-of-support integral that
respects the areal nature of each observation without evaluating the pairwise
spatial integrals that make full area-to-area kriging computationally
prohibitive --- reducing inference instead to standard sparse Bayesian linear
regression. Applied to a semi-synthetic albedo experiment with realistic OCO-2
footprint geometries, the framework achieves near-nominal coverage (95--98\%)
across model variants, with performance improving meaningfully when water land
cover is included as a covariate. Applied to real OCO-2 SIF observations, the
downscaled posterior maps align with NDVI and reflect the heterogeneity of the
urban landscape, with the strength of the SIF--NDVI relationship increasing
with observation density as expected from the uncertainty structure of the
model.

The framework is readily adaptable beyond the specific setting considered here.
The aggregation matrix $A$ encodes whatever footprint geometry is present,
accommodating regular or irregular supports; it is particularly well-suited to
instruments with irregular and overlapping footprints --- including
non-target-mode OCO-2, TROPOMI, or column-averaged CO$_2$ --- where existing
approaches are most strained. More broadly, the approach demonstrates that the
change-of-support problem for large collections of irregular areal observations
need not force a choice between geometric fidelity and computational
tractability: sparse discretization achieves both, while the Bayesian
formulation makes explicit what the data do and do not support at fine spatial
scales.

\bibliographystyle{abbrvnat}
\bibliography{reference}

@Misc{1, Title="{Keeping London moving}", Note = "\url{https://tfl.gov.uk/}", journal={Transport for London}}

@article{park2019overview,
  title={An overview of theoretical and practical issues in spatial downscaling of coarse resolution satellite-derived products},
  author={Park, No-Wook and Kim, Yeseul and Kwak, Geun-Ho},
  journal={Korean Journal of Remote Sensing},
  volume={35},
  number={4},
  pages={589--607},
  year={2019},
  publisher={The Korean Society of Remote Sensing}
}

@article{sun2018physics,
  title={A physics-based approach to oversample multi-satellite, multispecies observations to a common grid},
  author={Sun, Kang and Zhu, Lei and Cady-Pereira, Karen and Chan Miller, Christopher and Chance, Kelly and Clarisse, Lieven and Coheur, Pierre-Fran{\c{c}}ois and Gonz{\'a}lez Abad, Gonzalo and Huang, Guanyu and Liu, Xiong and others},
  journal={Atmospheric Measurement Techniques},
  volume={11},
  number={12},
  pages={6679--6701},
  year={2018},
  publisher={Copernicus GmbH}
}

@article{turner2020double,
  title={A double peak in the seasonality of California's photosynthesis as observed from space},
  author={Turner, Alexander J and K{\"o}hler, Philipp and Magney, Troy S and Frankenberg, Christian and Fung, Inez and Cohen, Ronald C},
  journal={Biogeosciences},
  volume={17},
  number={2},
  pages={405--422},
  year={2020},
  publisher={Copernicus GmbH}
}

@ARTICLE{Bonafoni,

  author={Bonafoni, Stefania and Sekertekin, Aliihsan},

  journal={IEEE Geoscience and Remote Sensing Letters}, 

  title={Albedo Retrieval From Sentinel-2 by New Narrow-to-Broadband Conversion Coefficients}, 

  year={2020},

  volume={17},

  number={9},

  pages={1618-1622},

  doi={10.1109/LGRS.2020.2967085}}

@article{hooten2014simultaneous,
  title={Simultaneous autoregressive (SAR) model},
  author={Hooten, Mevin B and Ver Hoef, Jay M and Hanks, Ephraim M},
  journal={Wiley StatsRef: Statistics Reference Online},
  pages={1--10},
  year={2014},
  publisher={John Wiley \& Sons, Ltd Chichester UK}
}

@article{Hodges01112010,
author = {James S. Hodges and Brian J. Reich},
title = {Adding Spatially-Correlated Errors Can Mess Up the Fixed Effect You Love},
journal = {The American Statistician},
volume = {64},
number = {4},
pages = {325--334},
year = {2010},
publisher = {ASA Website},
doi = {10.1198/tast.2010.10052},


URL = { 
    
        https://doi.org/10.1198/tast.2010.10052
    
    

},
eprint = { 
    
        https://doi.org/10.1198/tast.2010.10052
    
    

}

}

@article{GROSSMANN2018311,
title = {PhotoSpec: A new instrument to measure spatially distributed red and far-red Solar-Induced Chlorophyll Fluorescence},
journal = {Remote Sensing of Environment},
volume = {216},
pages = {311-327},
year = {2018},
issn = {0034-4257},
doi = {https://doi.org/10.1016/j.rse.2018.07.002},
url = {https://www.sciencedirect.com/science/article/pii/S0034425718303298},
author = {Katja Grossmann and Christian Frankenberg and Troy S. Magney and Stephen C. Hurlock and Ulrike Seibt and Jochen Stutz}
}

@article{li2018solar,
  title={Solar-induced chlorophyll fluorescence is strongly correlated with terrestrial photosynthesis for a wide variety of biomes: First global analysis based on OCO-2 and flux tower observations},
  author={Li, Xing and Xiao, Jingfeng and He, Binbin and Altaf Arain, M and Beringer, Jason and Desai, Ankur R and Emmel, Carmen and Hollinger, David Y and Krasnova, Alisa and Mammarella, Ivan and others},
  journal={Global change biology},
  volume={24},
  number={9},
  pages={3990--4008},
  year={2018},
  publisher={Wiley Online Library}
}

@article{eldering2017orbiting,
  title={The Orbiting Carbon Observatory-2 early science investigations of regional carbon dioxide fluxes},
  author={Eldering, A and Wennberg, PO and Crisp, D and Schimel, DS and Gunson, MR and Chatterjee, A and Liu, J and Schwandner, FM and Sun, Y and O’dell, CW and others},
  journal={Science},
  volume={358},
  number={6360},
  pages={eaam5745},
  year={2017},
  publisher={American Association for the Advancement of Science}
}

@article{zanaga2022esa,
  title={ESA WorldCover 10 m 2021 v200},
  author={Zanaga, Daniele and Van De Kerchove, Ruben and Daems, Dirk and De Keersmaecker, Wanda and Brockmann, Carsten and Kirches, Grit and Wevers, Jan and Cartus, Oliver and Santoro, Maurizio and Fritz, Steffen and others},
  year={2022},
  publisher={Zenodo}
}

@article{MOHAMMED2019111177,
title = {Remote sensing of solar-induced chlorophyll fluorescence (SIF) in vegetation: 50 years of progress},
journal = {Remote Sensing of Environment},
volume = {231},
pages = {111177},
year = {2019},
issn = {0034-4257},
doi = {https://doi.org/10.1016/j.rse.2019.04.030},
url = {https://www.sciencedirect.com/science/article/pii/S0034425719301816},
author = {Gina H. Mohammed and Roberto Colombo and Elizabeth M. Middleton and Uwe Rascher and Christiaan {van der Tol} and Ladislav Nedbal and Yves Goulas and Oscar Pérez-Priego and Alexander Damm and Michele Meroni and Joanna Joiner and Sergio Cogliati and Wouter Verhoef and Zbyněk Malenovský and Jean-Philippe Gastellu-Etchegorry and John R. Miller and Luis Guanter and Jose Moreno and Ismael Moya and Joseph A. Berry and Christian Frankenberg and Pablo J. Zarco-Tejada}
}

@article{calle2025generative,
  title={Generative multi-fidelity modeling and downscaling via spatial autoregressive transport maps},
  author={Calle-Saldarriaga, Alejandro and Wiemann, Paul FV and Katzfuss, Matthias},
  journal={arXiv preprint arXiv:2509.22474},
  year={2025}
}

@article{cheng2024recursive,
  title={Recursive nearest neighbor co-kriging models for big multi-fidelity spatial data sets},
  author={Cheng, Si and Konomi, Bledar A and Karagiannis, Georgios and Kang, Emily L},
  journal={Environmetrics},
  volume={35},
  number={4},
  pages={e2844},
  year={2024},
  publisher={Wiley Online Library}
}

@article{ma2022multifidelity,
  title={Multifidelity computer model emulation with high-dimensional output: An application to storm surge},
  author={Ma, Pulong and Karagiannis, Georgios and Konomi, Bledar A and Asher, Taylor G and Toro, Gabriel R and Cox, Andrew T},
  journal={Journal of the Royal Statistical Society Series C: Applied Statistics},
  volume={71},
  number={4},
  pages={861--883},
  year={2022},
  publisher={Oxford University Press}
}

@article{banerjee2017high,
  title={High-dimensional Bayesian geostatistics},
  author={Banerjee, Sudipto},
  journal={Bayesian analysis},
  volume={12},
  number={2},
  pages={583},
  year={2017}
}

@article{heaton2019case,
  title={A case study competition among methods for analyzing large spatial data},
  author={Heaton, Matthew J and Datta, Abhirup and Finley, Andrew O and Furrer, Reinhard and Guinness, Joseph and Guhaniyogi, Rajarshi and Gerber, Florian and Gramacy, Robert B and Hammerling, Dorit and Katzfuss, Matthias and others},
  journal={Journal of agricultural, biological and environmental Statistics},
  volume={24},
  number={3},
  pages={398--425},
  year={2019},
  publisher={Springer}
}

@article{zhang2018global,
  title={A global spatially contiguous solar-induced fluorescence (CSIF) dataset using neural networks},
  author={Zhang, Yao and Joiner, Joanna and Alemohammad, Seyed Hamed and Zhou, Sha and Gentine, Pierre},
  journal={Biogeosciences},
  volume={15},
  number={19},
  pages={5779--5800},
  year={2018},
  publisher={Copernicus GmbH}
}

@article{li2019global,
  title={A global, 0.05-degree product of solar-induced chlorophyll fluorescence derived from OCO-2, MODIS, and reanalysis data},
  author={Li, Xing and Xiao, Jingfeng},
  journal={Remote Sensing},
  volume={11},
  number={5},
  pages={517},
  year={2019},
  publisher={MDPI}
}

@article{gensheimer2022convolutional,
  title={A convolutional neural network for spatial downscaling of satellite-based solar-induced chlorophyll fluorescence (SIFnet)},
  author={Gensheimer, Johannes and Turner, Alexander J and K{\"o}hler, Philipp and Frankenberg, Christian and Chen, Jia},
  journal={Biogeosciences},
  volume={19},
  number={6},
  pages={1777--1793},
  year={2022},
  publisher={Copernicus Publications G{\"o}ttingen, Germany}
}

@article{DUVEILLER201672,
title = {Spatially downscaling sun-induced chlorophyll fluorescence leads to an improved temporal correlation with gross primary productivity},
journal = {Remote Sensing of Environment},
volume = {182},
pages = {72-89},
year = {2016},
issn = {0034-4257},
doi = {https://doi.org/10.1016/j.rse.2016.04.027},
url = {https://www.sciencedirect.com/science/article/pii/S0034425716301936},
author = {Gregory Duveiller and Alessandro Cescatti}
}

@article{jacobson2023spatial,
  title={Spatial statistical prediction of solar-induced chlorophyll fluorescence (SIF) from multivariate OCO-2 data},
  author={Jacobson, Josh and Cressie, Noel and Zammit-Mangion, Andrew},
  journal={Remote Sensing},
  volume={15},
  number={16},
  pages={4038},
  year={2023},
  publisher={MDPI}
}

@article{goovaerts2008kriging,
  title={Kriging and semivariogram deconvolution in the presence of irregular geographical units},
  author={Goovaerts, Pierre},
  journal={Mathematical geosciences},
  volume={40},
  number={1},
  pages={101--128},
  year={2008},
  publisher={Springer}
}

@misc{oco2siflite,
  author    = {{OCO-2 Science Team}},
  title     = {{OCO-2} {Level 2} Bias-Corrected Solar-Induced Fluorescence 
               and Other Select Fields from the {IMAP-DOAS} Algorithm 
               Aggregated as Daily Files, Retrospective Processing {V10r}
               ({OCO2\_L2\_Lite\_SIF\_10r})},
  year      = {2020},
  publisher = {NASA Goddard Earth Sciences Data and Information Services Center},
  doi       = {10.5067/XO2LBBNPO010}
}

@article{khan2022restricted,
  title={Restricted spatial regression methods: Implications for inference},
  author={Khan, Kori and Calder, Catherine A},
  journal={Journal of the American Statistical Association},
  volume={117},
  number={537},
  pages={482--494},
  year={2022},
  publisher={Taylor \& Francis}
}

@article{gilbert2025consistency,
  title={Consistency of common spatial estimators under spatial confounding},
  author={Gilbert, Brian and Ogburn, Elizabeth L and Datta, Abhirup},
  journal={Biometrika},
  volume={112},
  number={2},
  pages={asae070},
  year={2025},
  publisher={Oxford University Press}
}

@article{porcar2021chlorophyll,
  title={Chlorophyll a fluorescence illuminates a path connecting plant molecular biology to Earth-system science},
  author={Porcar-Castell, Albert and Malenovsk{\`y}, Zbyn{\v{e}}k and Magney, Troy and Van Wittenberghe, Shari and Fern{\'a}ndez-Mar{\'\i}n, Beatriz and Maignan, Fabienne and Zhang, Yongguang and Maseyk, Kadmiel and Atherton, Jon and Albert, Loren P and others},
  journal={Nature plants},
  volume={7},
  number={8},
  pages={998--1009},
  year={2021},
  publisher={Nature Publishing Group UK London}
}

@incollection{weaver2014processing,
  author    = {Weaver, Ron},
  title     = {Processing Levels},
  booktitle = {Encyclopedia of Remote Sensing},
  editor    = {Njoku, Eni G.},
  series    = {Encyclopedia of Earth Sciences Series},
  publisher = {Springer},
  address   = {New York, NY},
  year      = {2014},
  doi       = {10.1007/978-0-387-36699-9_36}
}

\section{Competing interests}
No competing interest is declared.

\section{Author contributions statement}

L.M. proposed the initial formulation of the aggregation matrix  and the idea of regularization.
L.C. proposed the use of a spatially correlated prior for regularization.
R.G. developed and formalized the methodology, implemented all software, carried out all experiments, and led the writing of the manuscript, including the refinement and analysis of the properties of the aggregation operator.
T.S.J., J.D., and L.R.H. contributed domain expertise, assisted with data acquisition and preparation, and provided feedback on the interpretation of results.
R.G., L.M., L.C., T.S.J., J.D., and L.R.H. reviewed and approved the final manuscript.

\section{Acknowledgments}
This work was supported by the National Aeronautics and Space Administration (NASA) under awards 80NSSC22K0161 and 80NSSC24K0752.

\end{document}